\documentclass{PoS}

\title{High-Performance I/O: HDF5 for Lattice QCD}

\ShortTitle{HDF5 for LQCD}

\author{Thorsten Kurth\\
Nuclear Science Division, Lawrence Berkeley National Laboratory, Berkeley, CA 94720, USA\\
Department of Physics, University of California, Berkeley, CA 94720, USA\\
E-mail: \email{tkurth@lbl.gov}}

\author{Andrew Pochinsky\\
Center for Theoretical Physics,
Massachusetts Institute of Technology, Cambridge, MA, USA\\
E-mail: \email{avp@mit.edu}}

\author{Abhinav Sarje\\
Computational Research Division, Lawrence Berkeley National Laboratory, Berkeley, CA 94720, USA\\
E-mail: \email{asarje@lbl.gov}}

\author{Sergey Syritsyn\\
RIKEN-BNL Research Center,
Brookhaven National Laboratory, Upton, NY 11973, USA\\
E-mail: \email{ssyritsyn@quark.phy.bnl.gov}}

\author{\speaker{Andr\'{e} Walker-Loud}\thanks{presented on behalf of Thorsten K\"{u}rth.}\\
Department of Physics, College of William and Mary, Williamsburg, Virginia 23187-8795, USA\\
Theory Group, Jefferson Laboratory, 12000 Jefferson Avenue, Newport News, VA 23606, USA\\
        E-mail: \email{walkloud@wm.edu}}

\abstract{
Practitioners of lattice QCD/QFT have been some of the primary pioneer users of the state-of-the-art high-performance-computing systems, and contribute towards the stress tests of such new machines as soon as they become available. As with all aspects of high-performance-computing, I/O is becoming an increasingly specialized component of these systems. In order to take advantage of the latest available high-performance I/O infrastructure, to ensure reliability and backwards compatibility of data files, and to help unify the data structures used in lattice codes, we have incorporated parallel HDF5 I/O into the SciDAC supported USQCD software stack. Here we present the design and implementation of this I/O framework. Our HDF5 implementation outperforms optimized QIO at the 10-20\% level and leaves room for further improvement by utilizing appropriate dataset chunking.}

\FullConference{The 32nd International Symposium on Lattice Field Theory,\\
		23-28 June, 2014\\
		Columbia University New York, NY}

\begin{document}

\section{Introduction}
The current era of peta-scale computing has brought lattice QCD (LQCD) to a mature state where the results of calculations are routinely compared with basic physical quantities measured experiments to both help understand the measurements as well as search for physics beyond the Standard Model, see for example ~\cite{Aoki:2013ldr}.
These advancements have led to efforts to solve increasingly complex problems ranging in application from relatively simple quantities needed for high-energy phenomenology to the more complex and computationally demanding quantities needed for basic nuclear physics.
As with all fields in need of high-performance-computing (HPC), data I/O is becoming an increasingly important aspect of the calculations.
Compared with other fields~\cite{SDAV:Byn2013a}, the I/O needs of LQCD are relatively mild.
However, the development of parallel file systems, such as \texttt{Lustre}, have lead to the need for increasingly specialized I/O software both to achieve good I/O performance and also to take advantage of the advanced user options available on these file systems.
In an effort to alleviate particular I/O bottlenecks LQCD is beginning to face, 
to prepare for future Exascale computing, 
and to encourage the use of a community wide professional I/O standard,
we were motivated to incorporate I/O software that:
is portable,
is standardized (i.e. third-party applications can read/write the files such as Mathematica, Python, MATLAB, ...),
is stable (IT-supported),
is not proprietary,
supports fast and reliable parallel I/O,
supports flexible data types
and simplifies data organization (a single file can contain configs, propagators, correlators, etc...).
We have found the Hierarchical Data Format v5 (HDF5)~\cite{hdf5} fulfills all these needs quite proficiently.

We report here on our implementation of parallel HDF5 into the USQCD Software Stack (supported by SciDAC~\cite{scidac} grants).  This implementation currently exists in QDP++ and Qlua~\cite{our_hdf5}.

\section{HDF5}
Hierarchical Data Format 5 (HDF5) is a \textit{technology suite designed to organize, store, discover, access, analyze, share and preserve diverse, complex data in continuously evolving heterogeneous computing and storage environments}~\cite{hdf5}.
HDF5 (and previous versions) is supported by the HDF Group with open-source software distributed at no cost.
HDF5 files are highly portable and can be accessed with many software interfaces such as MATLAB or Mathematica or via API libraries provided for programming languages such as C, C++, Java, Fortran90, Python, etc..
HDF files are self-describing, i.e. they contain information about endianness, dimensions of arrays and size of stored datatypes, floating point representation, etc.. They further allow users to specify complex data relationships and dependencies.
HDF5 supports many pre-defined datatypes as well as the creation of an unlimited variety of complex user-defined datatypes, allowing a single HDF file to contain many different data structures.
HDF files contain binary data stored naturally in n-dimensional arrays, where each element of the array may itself be a compound object, i.e. a color matrix, a propagator. These objects are themselves comprised of other compound objects, namely complex numbers.
The binary data is organized under a highly efficient metadata map (structured similar to a linux/unix directory) allowing for very fast access to portions of very large datasets.
HDF supports user defined metadata descriptions which can be stored as \textit{dataset/group attributes} or as separate \textit{datasets} in the data file.
Importantly, HDF also supports parallel I/O and is standard software at many HPC centers, often being used to stress test new parallel I/O file systems.
In other words, HDF provides a natural, efficient and subsidized home for lattice objects in a professionally maintained software environment that is accessible with standard scientific libraries and software.

\section{Implementation}
Many international research groups who perform calculations in lattice QCD use parts of the USQCD~\cite{usqcd} software stack~\cite{usqcd-software}.
The stack is comprised of four layers: the bottom layer contains three packages containing optimized linear algebra (QLA) and parallelization routines (QMP, QMT).  The second layer contains software packages which provide the most basic functionality necessary for performing LQCD calculations (QDP/QDP++) including a variety of data types for describing lattice objects such as gauge configurations, quark propagators, etc.
This second layer also contains the QIO package which is designed to provide fast parallel I/O of lattice objects.
The frameworks we have focussed on so far are QDP++ and Qlua~\cite{qlua}.  We also have an initial interface to the HDF5 extension of QDP++ in \texttt{Chroma}~\cite{Edwards:2004sx}.

\subsection{QDP++}
Inspired by the commonly used \texttt{XMLWriter/XMLReader} interface of QDP++, we implemented the classes \texttt{HDF5Writer} and \texttt{HDF5Reader}. These classes act as a C++ wrapper for the C HDF5 interface. Their base class HDF5 cannot be instantiated by the user, but contains members common to both children, such as closing the current file, reading data, traversing the group hierarchy, etc. The \texttt{HDF5Writer} provides additional capabilities such as writing to a file, deleting objects, etc. The user can also create groups and attach attributes to objects, and thus can freely organize the content of the file. The syntax resembles that of the \texttt{XMLWriter/XMLReader} discussed above, but it also contains additional features: the user can navigate through the group hierarchy by using the member function \texttt{cd(path)}, which resembles the corresponding UNIX command. The current working directory is returned by \texttt{pwd()}. Functions which resemble other useful UNIX commands such as \texttt{ls(path)} will be added in future releases.

The most important aspect of our implementation is reliable and performant parallel I/O, especially of large lattice objects. 
The lattice sites are always arranged into lexicographical order: the components of gauge configurations and lattice quark propagators are thus ordered (from slowest to fastest) $[t,z,y,x,\mu,c_2,c_1]$ and $[t,z,y,x,s_2,s_1,c_2,c_1]$
 respectively. Here, the $c_i,s_i$ are the color and spin components respectively and $\mu$ is the direction of the corresponding link. Note that the memory layout in QDP/QDP++ for a gauge configuration is $[\mu,t,z,y,x,c_2,c_1]$, hence the directional index $\mu$ is the slowest. We reorder the data since different $\mu$-components reside on the same node allowing for I/O operations to be performed on chunked rather than on strided data, gaining a significant speedup in performance. By default, datasets are not stored in \texttt{H5D\textunderscore CHUNKED} layout, however, the user can enable chunking by calling the \texttt{set\textunderscore stripesize} member function of \texttt{HDF5Writer} with a non-negative integer number in the argument representing the number of chunks.

An important aspect of our HDF5 implementation is data integrity. All objects are closed as soon as they are not needed any more: datasets and attributes are closed immediately after reading from or writing into them. Only the current group (that returned by \texttt{pwd()}) as well as all its parent groups are kept open until the file is closed. However, any time a \texttt{cd(path)} command is issued, it is checked if open groups can be closed. This is the case if the user changes to a location higher in the tree or a different branch of the tree. This behavior minimizes data loss in case the main QCD application crashes before the HDF5 file was closed, as HDF5 protects the closed objects from data corruption.

\subsection{QLUA}

The Qlua~\cite{qlua} interface to HDF5 follows the narrow interface design
approach used for Qlua's QIO. In Qlua, there are separate HDF5 writer
and reader objects used for output and input respectively. This
reflects the I/O pattern found in LQCD applications and allows the
implementation to be optimized separately for read and write
operations.

When a HDF5 writer is opened or created,
the user can specify global file options (e.g., which HDF5 file driver
to use, how large chunks should be, and some other HDF5 parameters) as
well as the default options for writing objects into the file. All
write operations are dispatched from a single procedure,
\texttt{hf:write(path, object[, options])}. The write procedure decides
on data representation depending on the type of the
\texttt{object}. Components of \texttt{path} are created in the file
as needed. 
The datatype of the data element is computed and looked up
in the file; if this datatype was not used before, it is stored in the
file, otherwise, a clone of the existing datatype is used for this
object. One can overwrite the default options by providing an
\texttt{options} table. 
All options, including the width of floating
point can be controlled this way. Data is always written in the
big-endian byte order.

For each stored object, a checksum is computed and stored in an
object's attribute \texttt{sha256}. For sequential data, this is the
standard SHA-2 checksum \cite{shs2012}. For parallel data, checksums
are computed on each lattice site from the site data and lattice
coordinates and are XORed together. This approach allows one to
compute the checksum in parallel; the result is believed to provide
sufficient collision resistance.

For each data object, the value of the \texttt{kind} attribute is
determined by the type of the object. Qlua also writes a timestamp in
microseconds since the UNIX epoch into \texttt{time} attribute.

The Qlua HDF5 reader by default expects to find \texttt{kind} and
\texttt{sha256} attributes. It uses the value of the \texttt{kind}
attribute to select an appropriate reader for the object, and checks bits
read from storage against the stored checksum to detect corrupted
data. It is possible to read non-conformant data by overwriting
\texttt{kind}. The data integrity check can be disabled if
needed. Both kind and checksum can be controlled on per-object level.

The library also provides a set of convenience methods,
\texttt{h:ls(path)} returns a list of object names in a group,
\texttt{h:stat(path)} returns a table of attributes and the type of
the object, and \texttt{h:chdir(path)} changes the current location in
the HDF5 file. In all calls the \texttt{path} parameter is interpreted
as starting from the root group if \texttt{path} begins with a slash \texttt{"/"},
or as the relative path to the current location otherwise. The current location
can be queried by calling \texttt{h:cwd()}.

\section{Performance}
We report on a few tests performed of the HDF5 implementation on several top ranking supercomputers: Hopper, a Cray XE6 system at NERSC; Edison, an XC30 system at NERSC and Titan, a Cray Xk7 system at OLCF.
The tests were performed with lattice configurations with volumes ranging from $V=32^3\times48$ to $V=128^3\times256$.

\subsection{HDF5 vs QIO}
\begin{figure}[]
\centering
\includegraphics[width=0.6\textwidth]{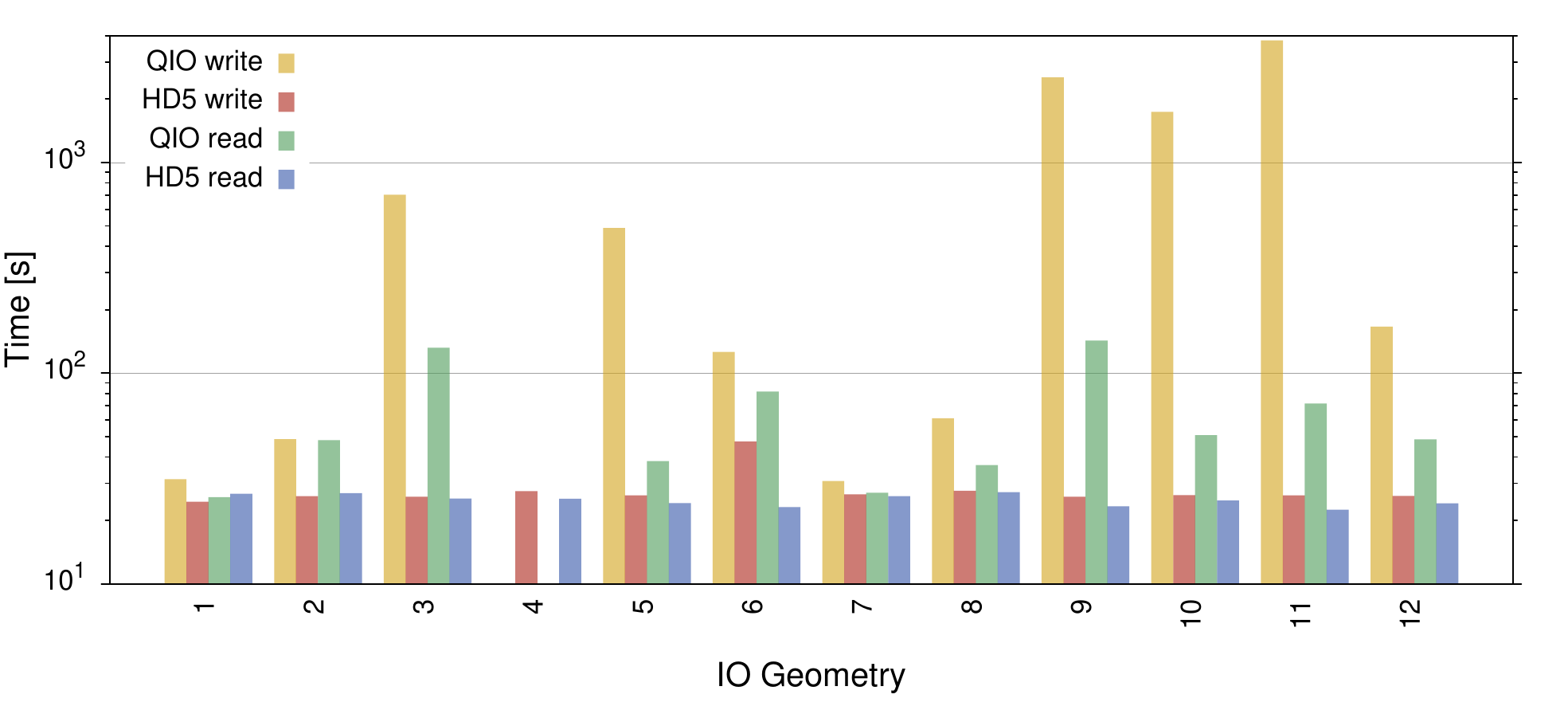}
\includegraphics[width=0.6\textwidth]{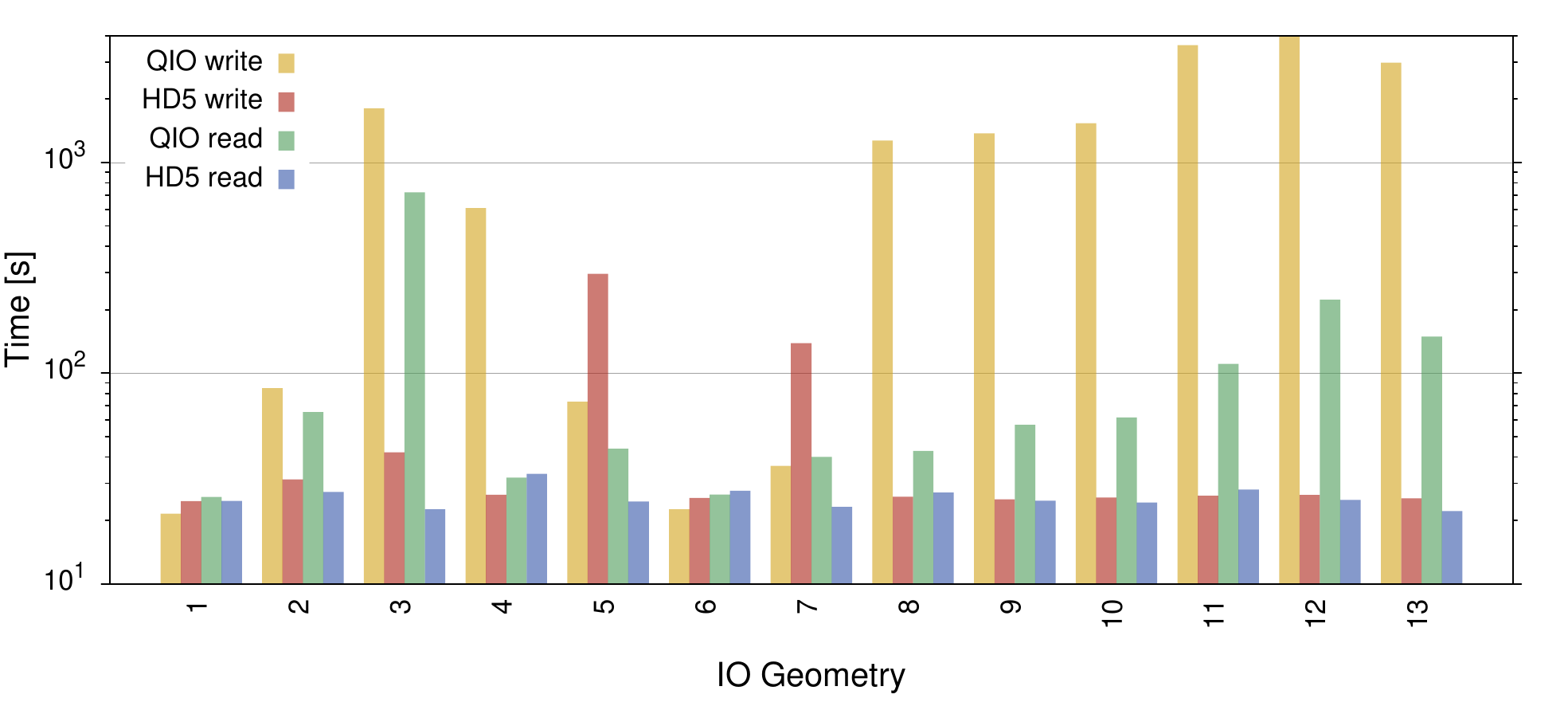}
\caption{The performance with different I/O geometries. Each I/O geometry represents a different configuration of the number of I/O node distributed along the four dimensions of the lattice. These are for a lattice of size $64\times 64\times 64\times\ 128$, using 384 core (top) and 3,072 cores (bottom), with one I/O process per node (24 cores). Examples of the geometries for the first case (16 I/O nodes) are, $1: 1^3\times 16, 2: 1\times 1\times 16\times 1, 5: 2\times 1\times 1\times 8, 10: 2^4$, and so on.}
\label{fig:iogeom}
\end{figure}

We optimized the performance of QIO, the I/O driver in the USQCD software stack, by utilizing the \texttt{-iogeom} flag on a \texttt{striped Lustre} file system.  We found that the HDF5 implementation tended to be both faster and more stable, see Figure~\ref{fig:iogeom}.

\subsection{Scaling}
\begin{figure}[]
\centering
\includegraphics[width=0.48\textwidth]{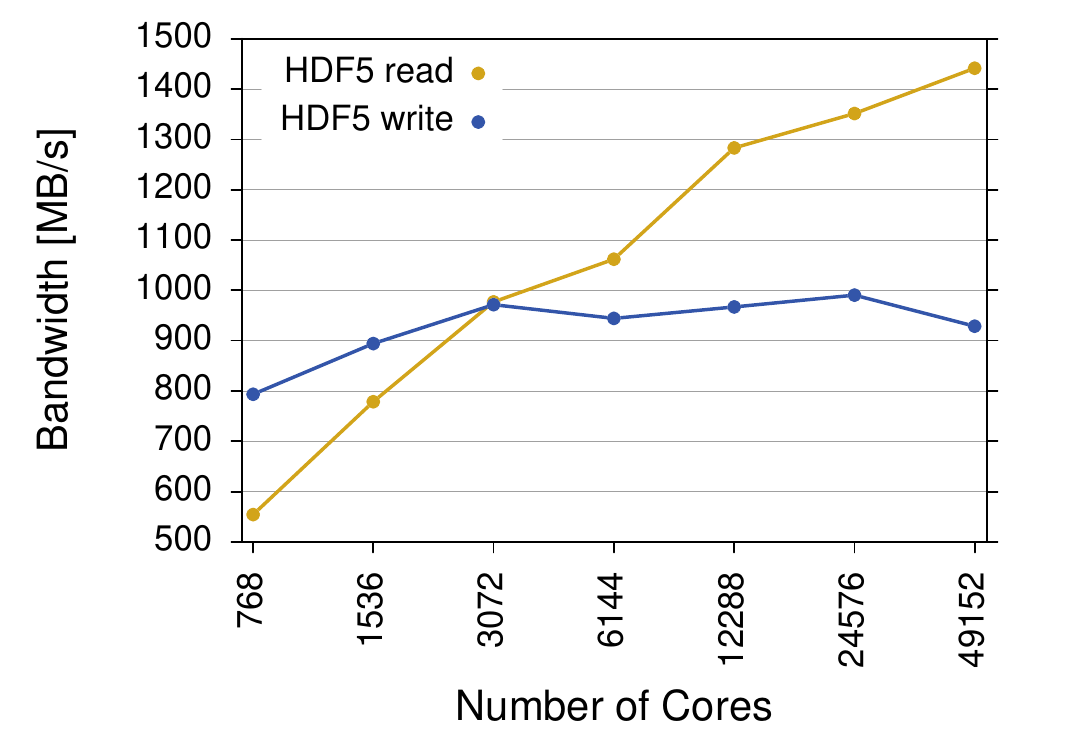}
\includegraphics[width=0.48\textwidth]{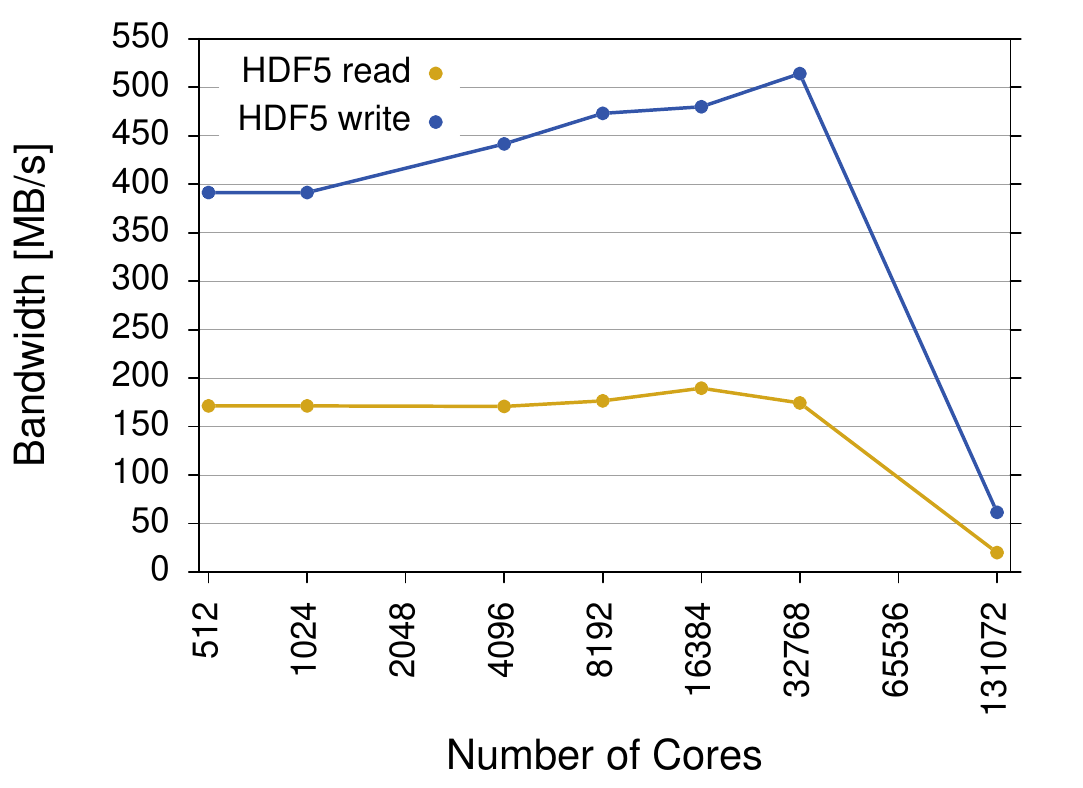}
\includegraphics[width=0.48\textwidth]{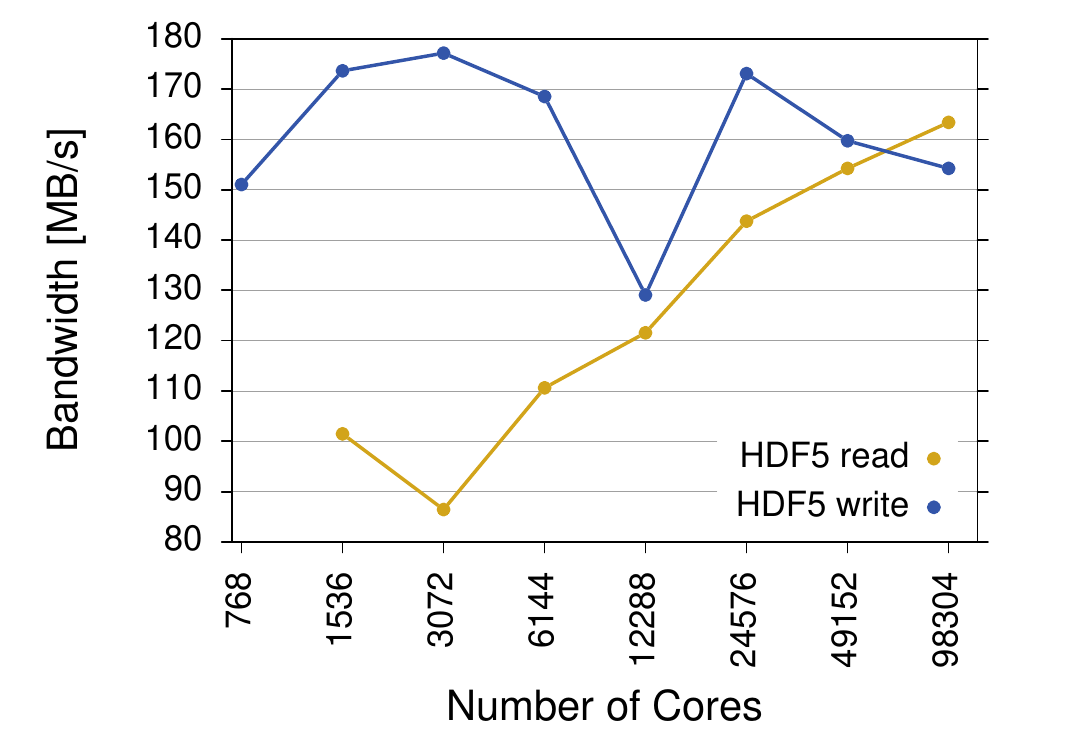}
\includegraphics[width=0.48\textwidth]{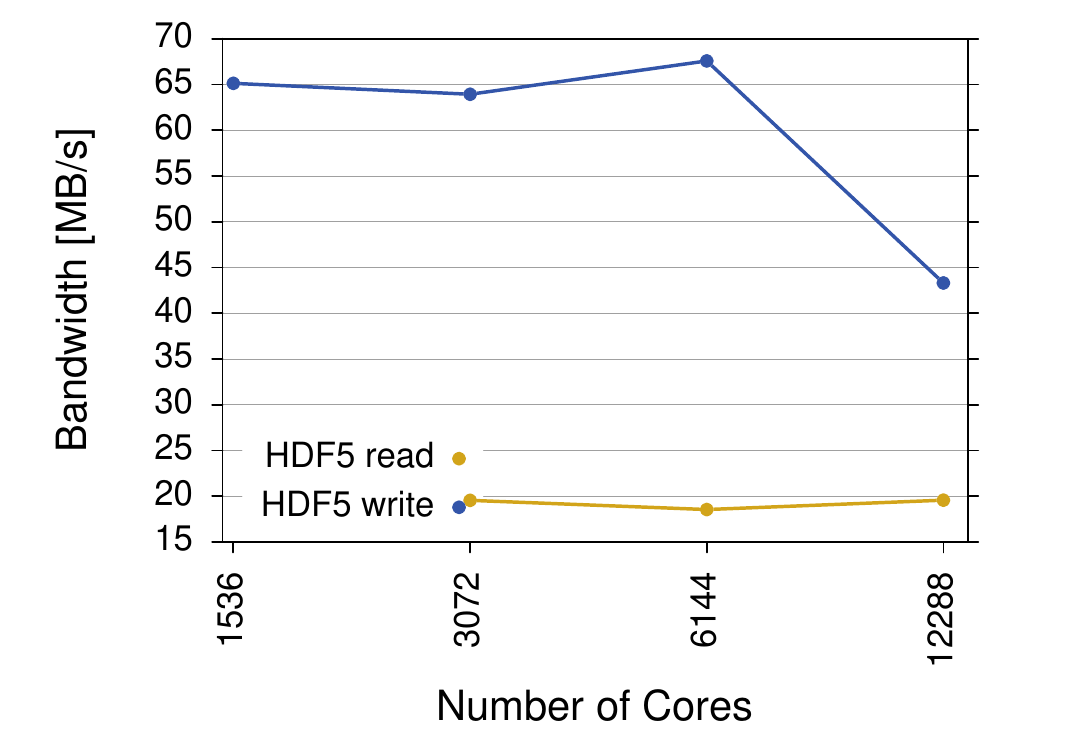}
\caption{HDF5 I/O weak- (top) and strong-scaling (bottom) on Edison (left) and Titan (right) on a $128^3\times256$ lattice.}
\label{fig:scaling}
\end{figure}

We tested the weak and strong scaling of the HDF5 I/O on two architectures, Edison and Titan, see Figure~\ref{fig:scaling}.

\subsection{Lustre Striping}
\begin{figure}[]
\centering
\includegraphics[width=0.6\textwidth,height=3.8cm]{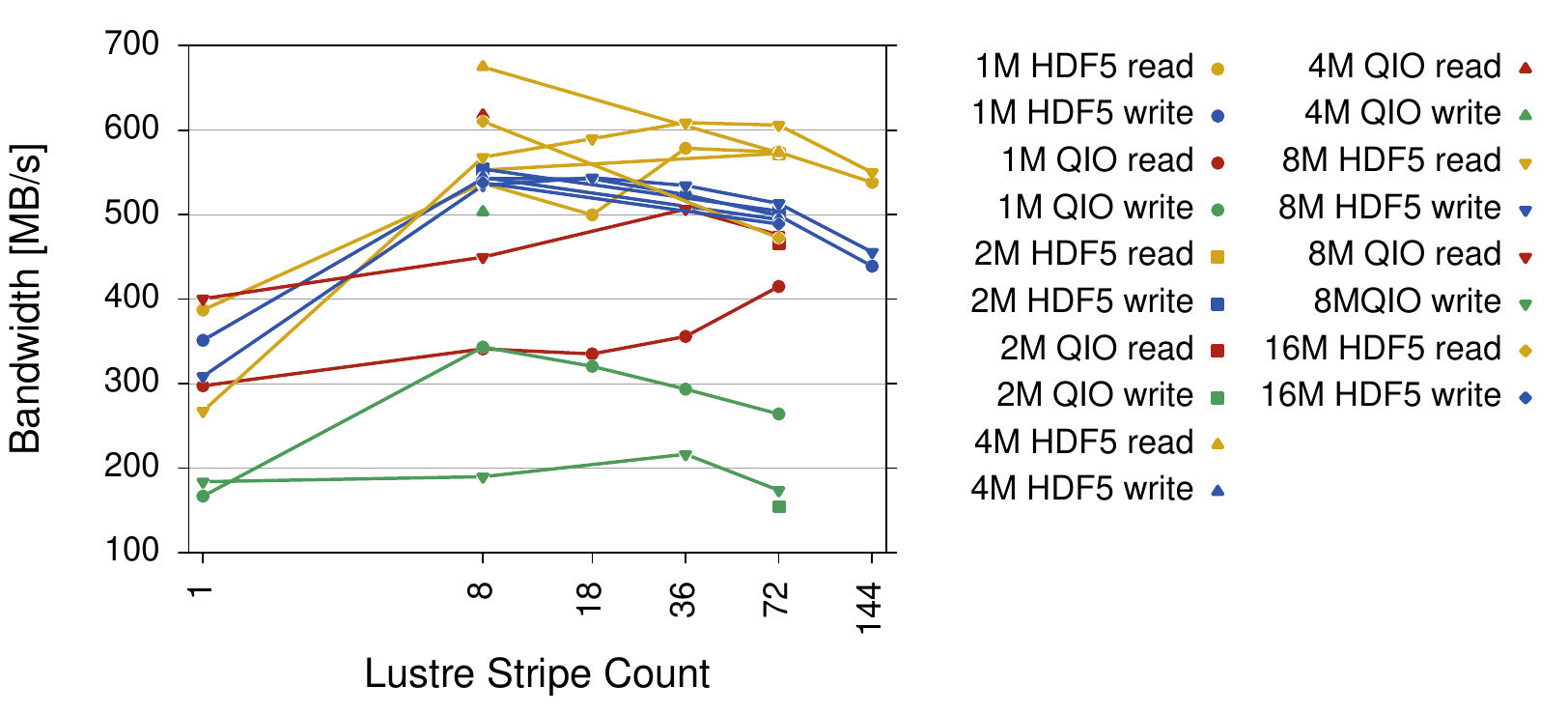}
\includegraphics[width=0.48\textwidth,height=3.8cm]{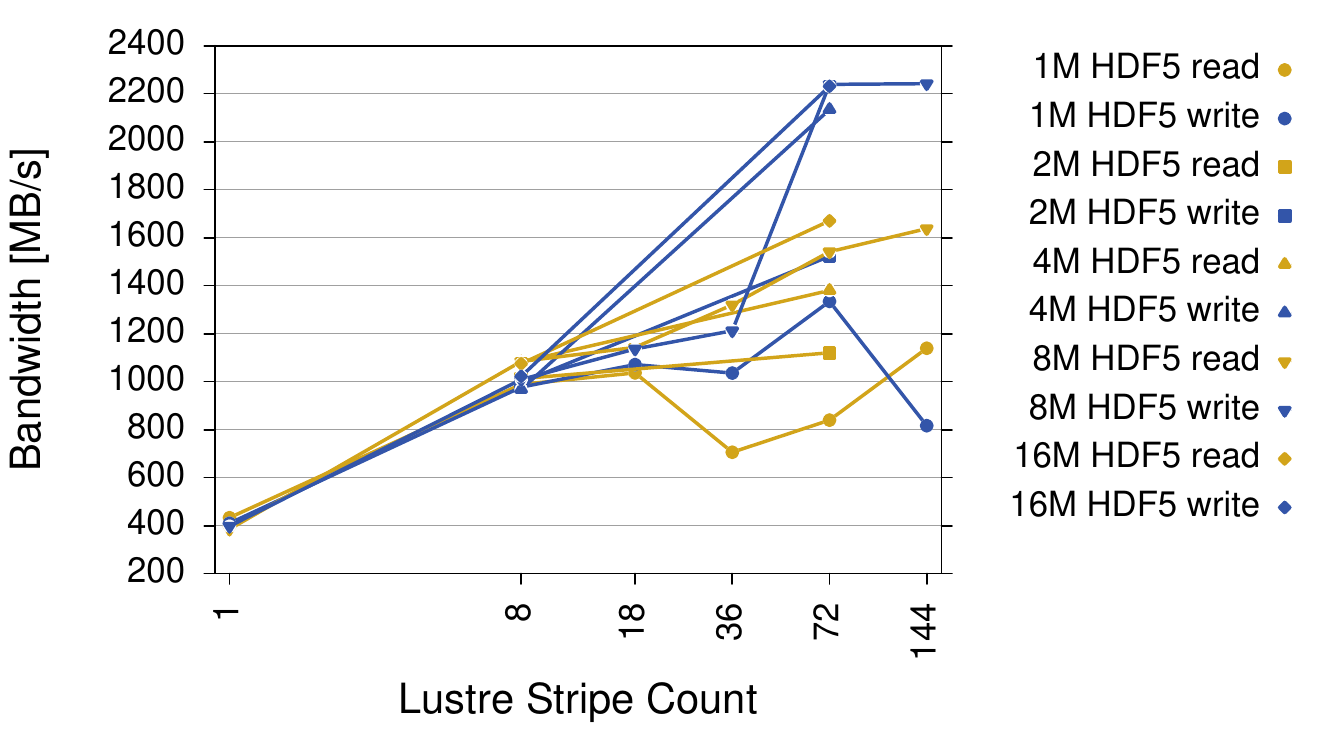}
\includegraphics[width=0.48\textwidth,height=3.8cm]{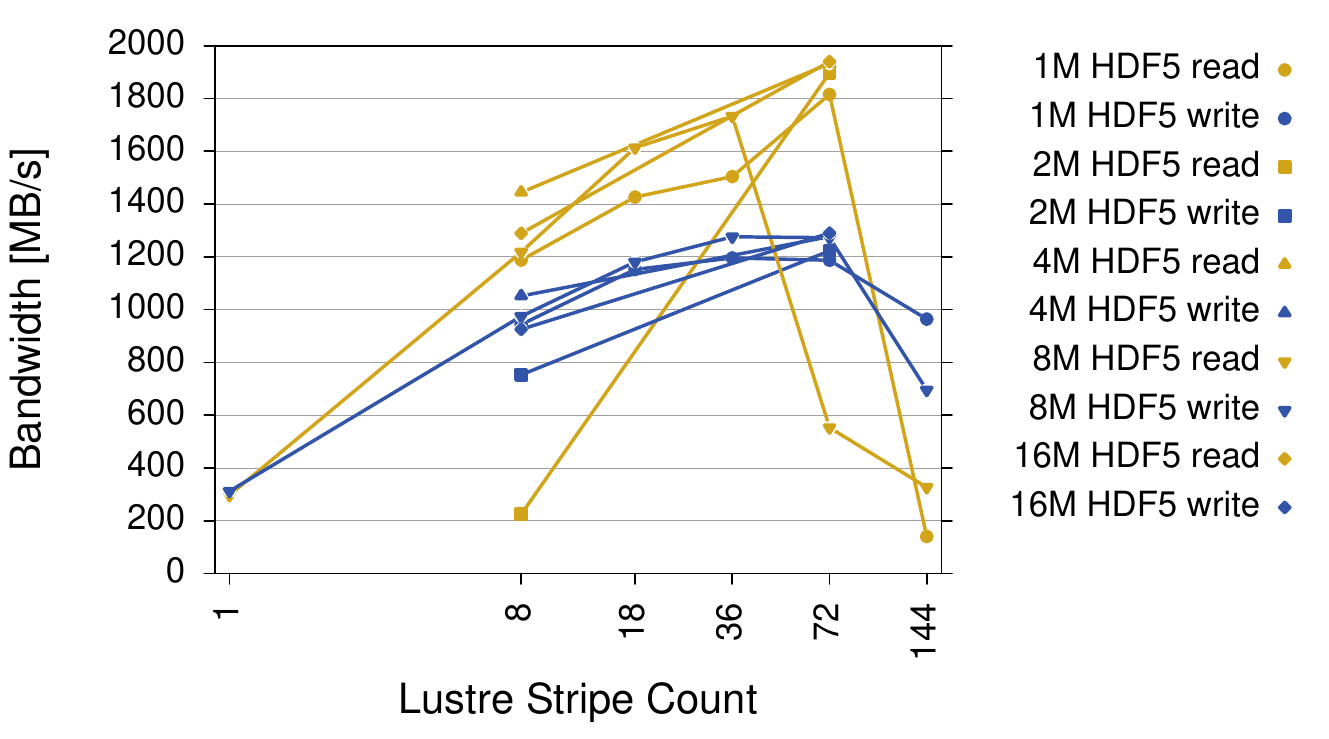}
\includegraphics[width=0.48\textwidth]{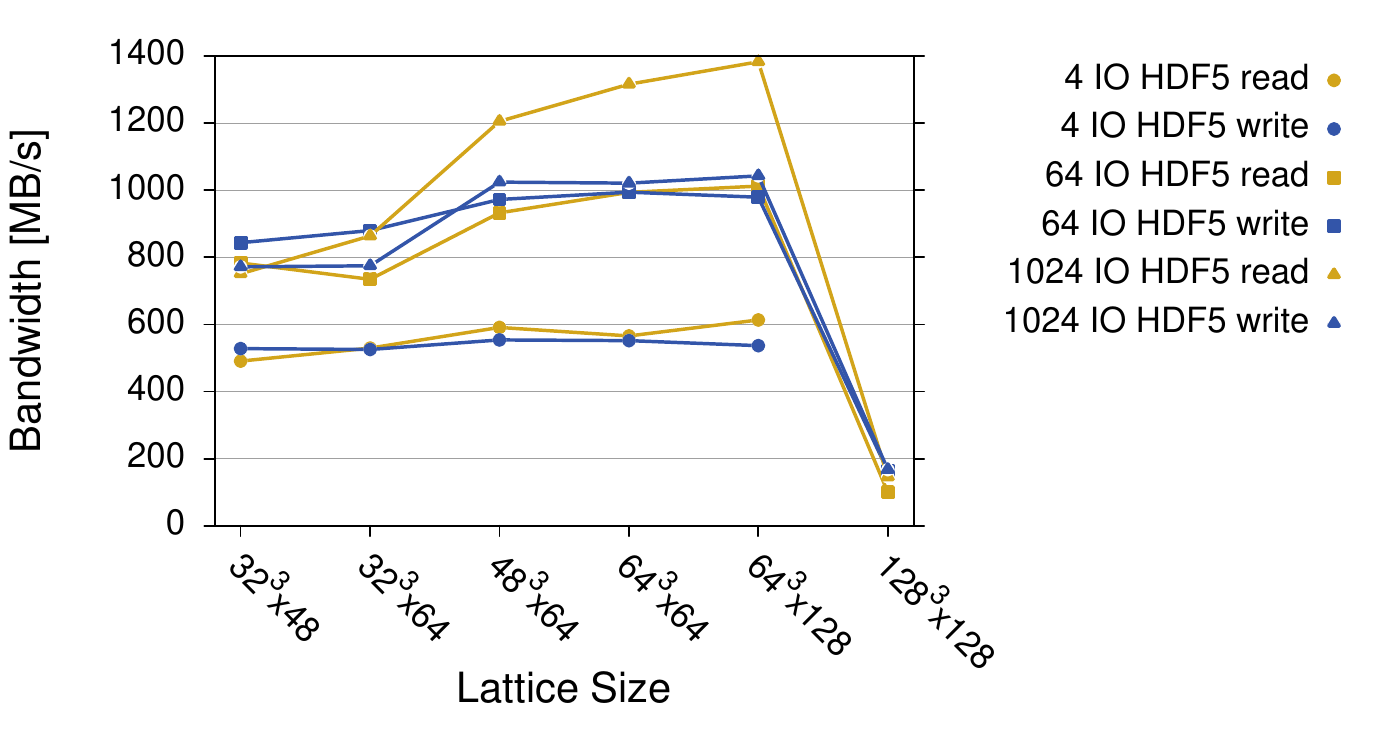}
\includegraphics[width=0.48\textwidth]{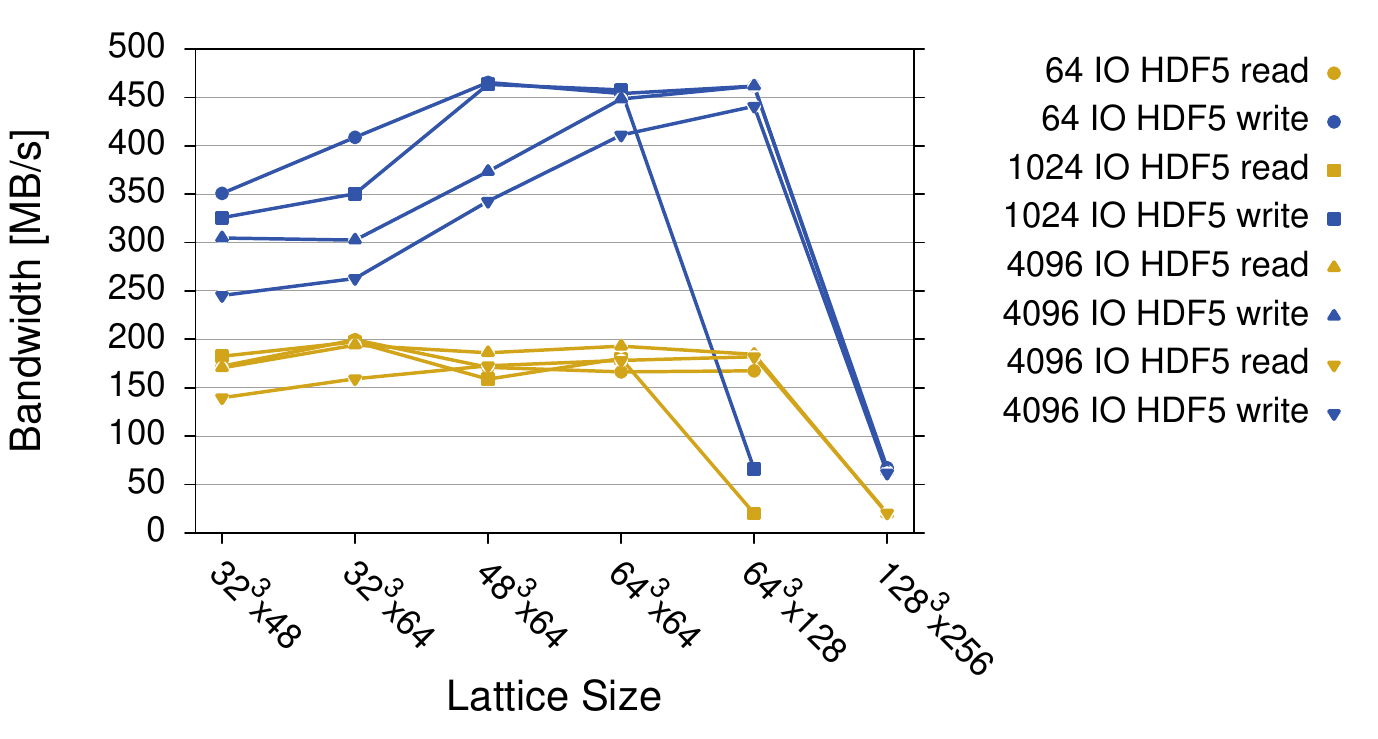}
\caption{Performance dependence on stripe count and size for 4 (top), 64(middle-left) and 1024 I/O nodes (middle-right) on Edison, and for 4 and 64 I/O nodes on Edison (bottom-left) and Titan (bottom-right).}
\label{fig:striping}
\end{figure}

The I/O performance can be significantly influenced by the size and number of stripes as well as the chunk size of the datasets, see Figure~\ref{fig:striping}.

\section{Conclusions}
The I/O needs for LQCD have grown significantly in recent years and is now becoming an important issue to deal with, as it is for all HPC consumers.
Instead of continuing to modify our existing QIO software, a task requiring increasing time and effort, we have opted to explore 3rd party software to handle our I/O needs.
The HDF Group has decades of experience dealing with HPC parallel I/O and has developed a first rate software suite which is non-proprietary, highly portable and easy to interface with standard scientific libraries.
To ease the hopeful transition to using HDF5, we have modeled the \texttt{HDF5Writer} and \texttt{HDF5Reader} classes after the QDP++ \texttt{XMLWriter/XMLReader} interface.  A more significant LQCD community use of HDF5 will also aid in forming more unified data structures aiding in the sharing of configurations and other data files on the ILDG.

\section*{Acknowledgements}

TK, AS and AWL acknowledge many helpful conversations with B\'{a}lint Jo\'{o}.  AS would also like to thank Samuel Williams for input.  This work was supported in part from the DOE SciDAC-3 grants for the CalLat Collaboration (TK, AS, SS, AWL) and the USQCD Collaboration (AP).
Authors from Lawrence Berkeley National Laboratory were supported by the DOE Office of Advanced Scientific Computing Research under contract number DE-AC02-05CH11231 (AS). This work was also in part funded by the U.S. Department of Energy Office of Nuclear Physics under grant number DE-FG02-94ER40818 (AP).

\end{document}